\documentclass[10pt]{article}
\usepackage[OE]{express}

\begin{document}
\title{Enhancing performance of coherent OTDR systems with polarization diversity complementary codes}

\author{Christian Dorize\authormark{1,2} and Elie Awwad\authormark{1,3}}

\address{\authormark{1}Nokia Bell Labs Paris-Saclay, 1 route de Villejust, 91620 Nozay, FRANCE\\
\authormark{2}christian.dorize@nokia-bell-labs.com\\
\authormark{3}elie.awwad@nokia-bell-labs.com}




\begin{abstract}
Monitoring the optical phase change in a fiber enables a wide range of applications where fast phase variations are induced by acoustic signals or vibrations in general. However, the quality of the estimated fiber response strongly depends on the method used to modulate the light sent to the fiber and capture the variations of the optical field. In this paper, we show that distributed optical fiber sensing systems can advantageously exploit techniques from the telecommunication domain, as those used in coherent optical transmission, to enhance their performance in detecting mechanical events, while jointly offering a simpler setup than widespread pulse-cloning or spectral-sweep based schemes with acousto-optic modulators. We periodically capture an overall fiber Jones matrix estimate thanks to a novel probing technique using two mutually orthogonal complementary (Golay) pairs of binary sequences applied simultaneously in phase and quadrature on two orthogonal polarization states. A perfect channel response estimation of the sensor array is achieved, subject to conditions detailed in the paper, thus enhancing the sensitivity and bandwidth of coherent $\phi$-OTDR systems. High sensitivity, linear response, and bandwidth coverage up to $18~\mathrm{kHz}$ are demonstrated with a sensor array composed of 10 fiber Bragg gratings (FBGs).
\end{abstract}



\section{Introduction}

Fiber optic sensors, being intrinsically immune to electromagnetic interference and fairly resistant in harsh environments, meet a growing interest in monitoring applications (structural health monitoring, railway surveillance, pipeline monitoring...). Distributed fiber optic sensors based on optical reflectometry make use of a variety of light scattering effects occurring in the fiber such as Raman, Brillouin, and Rayleigh backscattering to measure temperature (with any of the three effects) or mechanical variations such as strains (only with the two latter)~\cite{Mas16}. Optical fiber sensors may also be customized or enhanced by periodically inscribing fiber Bragg gratings (FBGs) to amplify the backscattered optical field~\cite{Mas16} resulting in a quasi-distributed system with a resolution fixed by the distance between gratings. The main characteristics of a distributed sensor are its sensitivity, spatial resolution and maximum reach. Another important feature for dynamic phenomena distributed sensing is the bandwidth of the mechanical events that the sensor is able to detect, which is closely related to the targeted sensitivity and the sensor length.
 
\textit{Detecting} and \textit{quantifying} sound waves and vibrations, known as distributed acoustic sensing (DAS) or distributed vibration sensing (DVS) is critical in areas of geophysical sciences and surveillance of sensitive sites or infrastructures. Phase and coherent optical-time-domain (resp. optical-frequency-domain) reflectometry (OTDR, resp. OFDR) systems are usually based on an interrogator sending one or more short light pulses or frequency sweeps~\cite{Pal13,Shi15,Yan16,Fan17}. The detector consists of a simple photodiode if, for instance, two pulses at slightly different frequencies are separately launched in the sensing fiber~\cite{Mas16}. In case single pulses are sent, an imbalanced Mach-Zehnder interferometer and a phase detector, or a balanced coherent detector that mixes the backscattered pulse with a local oscillator are used at the receiver side to detect relative phase changes in the Rayleigh backscattered optical field~\cite{Mas16,Shi15,Yan16,Fan17}. The main limitations of these phase-OTDR systems are firstly a trade-off between the spatial resolution and the maximum reach, given that a high spatial resolution forces the use of short pulses resulting in a low signal-to-noise ratio, secondly a trade-off between maximum reach and the covered mechanical bandwidth, the latter being equal to half of the scanning rate of the pulses. A reflectometry scheme based on the injection of several linear-frequency-modulated probe pulses was suggested in~\cite{Che17} to relax these two trade-offs, showcasing a $9~\mathrm{kHz}$ bandwidth with a $10~\mathrm{m}$ resolution over a $24.7~\mathrm{km}$-long fiber. However, the interrogators in these schemes all rely on individual probing pulses generated by acousto-optic modulators or even more complex structures. They are also vulnerable to polarization fading effects given that the Rayleigh backscattered light is polarization dependent. A dual-polarization coherent receiver allowing to detect all the backscattered information by projecting the received optical field over two orthogonal polarization states can fix this problem as shown in recent works~\cite{Mar16,Yan17}. 
In order to further relax the reach-spatial resolution trade-off, our approach in this paper consists in continuously probing the sensor using a training sequence that modulates the optical carrier injected in the fiber, as done in~\cite{Mar16}. While random binary sequences modulate two polarization states to probe a $500~\mathrm{m}$-long sensor and detect a sinusoidal strain of $500~\mathrm{Hz}$ in~\cite{Mar16}, a perfect optical channel estimation can only be reached asymptotically for very large sequences. Hence, we design in this work optimized probing sequences of finite length allowing to extend the covered bandwidth. The proposed DAS scheme consists in transmitting polarization-multiplexed coded sequences designed from complementary Golay pairs, and detecting the backscattered optical signal using a polarization-diversity coherent receiver typically used in optical fiber transmission systems~\cite{Kik16} followed by a correlation-based post-processing to extract the channel response. As is well known, Rayleigh backscattering is randomly distributed along the fiber and the distributed scatterers reflect different amounts of energy. For this reason, in order to concentrate on the performance of the proposed interrogator, the experimental part of this paper focuses on a fiber sensor with explicit and deterministic back-reflectors using periodically inserted FBGs that turn the fiber into a sensor array, as in~\cite{Zhu15,Sun17}, with a resolution of $10~\mathrm{m}$. We show that the proposed DAS solution is capable of spatially resolving up to $18~\mathrm{kHz}$ dynamic strains even after displacing the sensor array by $25~\mathrm{km}$ of SMF. 
 
The paper is organized as follows: in section 2, we introduce the theory underpinning the coded sequences designed to scan the sensor array through polarization multiplexing; in section 3, we describe the experimental setup built to test the DAS system; the results are given in section 4 in static mode first to quantify the noise limits, followed by a dynamic mode analysis during which the sensor array is perturbed at two different locations by two independent vibrations.    


\section{Theory}

\subsection{Notation}
In the following, $\ast$ and $\otimes$ operators denote the convolution and the correlation operators respectively; $\delta(t)$ stands for the delta function. The correlation and convolution between signals $a(n)$ and $b(n)$ are related since $a(n)\otimes b(n) = a(n)\ast b^\ast(-n)$, $b^\ast$ standing for the conjugated complex of $b$ and $n$ being a time index. $E_{tx}$ and $E_{ty}$ (resp. $E_{rx}$ and $E_{ry}$) denote the two polarization tributaries of the optical field at the transmitter side (resp. receiver side). The optical field vector generated at the transmitter side is given by:
\begin{equation}
\overrightarrow{E}_t (n) = \left[ \begin{array}{c}
   A_{tx}(n)\exp(i\phi_{tx}(n)) \\
   A_{ty}(n)\exp(i\phi_{ty}(n)) \\
  \end{array}\right]\exp(i2\pi\nu_0nT_S +\phi_0(n)),~~~~n=[1\ldots N]
\end{equation}
where $A_{tx}$, $A_{ty}$ are the modulated amplitudes of the x- and y- polarization tributaries, $\phi_{tx}$, $\phi_{ty}$ are the modulated phases of the x- and y- polarization tributaries,$\nu_0$ is the optical carrier frequency, $\phi_0$ is the phase noise generated by the laser and $T_S$ the symbol duration.

The impulse response of a fiber section is represented by a $2\times2$ Jones matrix:
\begin{equation}
\mathbf{H} = \left[ \begin{array}{cc}
   h_{xx}~~ h_{xy}\\
   h_{yx}~~ h_{yy}\\
  \end{array}\right]
\end{equation}
where $h_{xx,xy,yx,yy}$ are complex numbers describing the relation between the polarization tributaries at the input and output of a fiber section. The location and characterization of any mechanical excitation impacting the sensor array is extracted from the space-time table of impulse responses periodically estimated at each fiber array section. 

\subsection{Design of polarization-diversity coded sequences}
Our objective is twofold: achieving a perfect estimate of the Jones matrix impulse response and maximizing the number of estimates per time unit to enhance the covered mechanical bandwidth. 
Let us consider two real binary sequences of size $N_G$ each that form a complementary, or Golay, pair~\cite{Gol61} such as:
\begin{equation}
G_{a1}(n)\otimes G_{a1}(n) + G_{b1}(n)\otimes G_{b1}(n) =  \delta(n)
\label{Gol1}
\end{equation}
Thanks to the above complementary property, probing a channel with such a sequence pair allows for a perfect impulse response estimation in case of a basic single-input-single-output channel. Practically, the transmission of the two complementary sequences is applied successively in time and the response estimation is extracted after a correlation-based post-processing at the receiver side~\cite{Naz89}. Notice that perfect estimation requires that the channel remains stationary during the overall probing time.

A natural extension of the single-input-single-output channel case to the $2\times2$ Jones matrix impulse response consists in successively probing each of the two polarization tributaries by means of the above procedure. However, it still extends the probing time, thus reducing the number of impulse response estimations per second and impacting the system bandwidth. 

Today's optical transmission systems, based on coherent technology, use a polarization diversity transmitter and receiver to jointly propagate independent signals onto each of the two orthogonal polarization axes. This polarization degree of freedom is generally underused in the fiber sensing domain. To the authors knowledge, the sole work that considered polarization diversity at the transmitter and at the receiver side is~\cite{Mar16}. Our purpose is to study, for complementary codes, the conditions to achieve a perfect Jones matrix estimation with a simultaneous probing of the two polarization axes, thus keeping the channel stationarity constraint the same as for a single-input-single-output channel.



During a period of $N$ symbol times, we modulate the $x$ (resp. $y$) polarization of the optical signal at the transmitter side by $N_G$-long sequences $G_x(n) = G_{xI}(n)+i G_{xQ}(n)$ (resp. $G_y(n) = G_{yI}(n)+i G_{yQ}(n)$) at a given symbol-rate $F_S=1/T_S$, and send zeros during the remaining $N-N_G$ slots. Hence:
\begin{equation}
E_{tx,ty}(n+kN) = \begin{cases}
~~G_{xI,yI}(n)+i G_{xQ,yQ}(n) & \text{if}~ n~\text{mod}~N\leq N_G\\
~~0&\text{elsewhere} 
\end{cases}
\end{equation}
Let $E_{rx}(n)$ and $E_{ry}(n)$ be the sampled outputs of a coherent polarization diversity receiver at a rate of one sample per symbol. They are given by the convolution of the transmitted signal and the impulse response of the sensor array:
\begin{equation}
\begin{split}
E_{rx}(n) &= h_{xx}(n)\ast E_{tx}(n)+h_{xy}(n)\ast E_{ty}(n)\\
E_{ry}(n) &= h_{yx}(n)\ast E_{tx}(n)+h_{yy}(n)\ast E_{ty}(n)
\end{split}
\end{equation} 
In the following, we only develop $E_{rx}(n)$ for sake of simplicity since a similar procedure can be applied to $E_{ry}(n)$. 

\begin{figure}[htbp]
\centering
\begin{picture}(400,170)
\put(0,0)
{
\put(-10,5){\includegraphics[width=220pt,height=170pt]{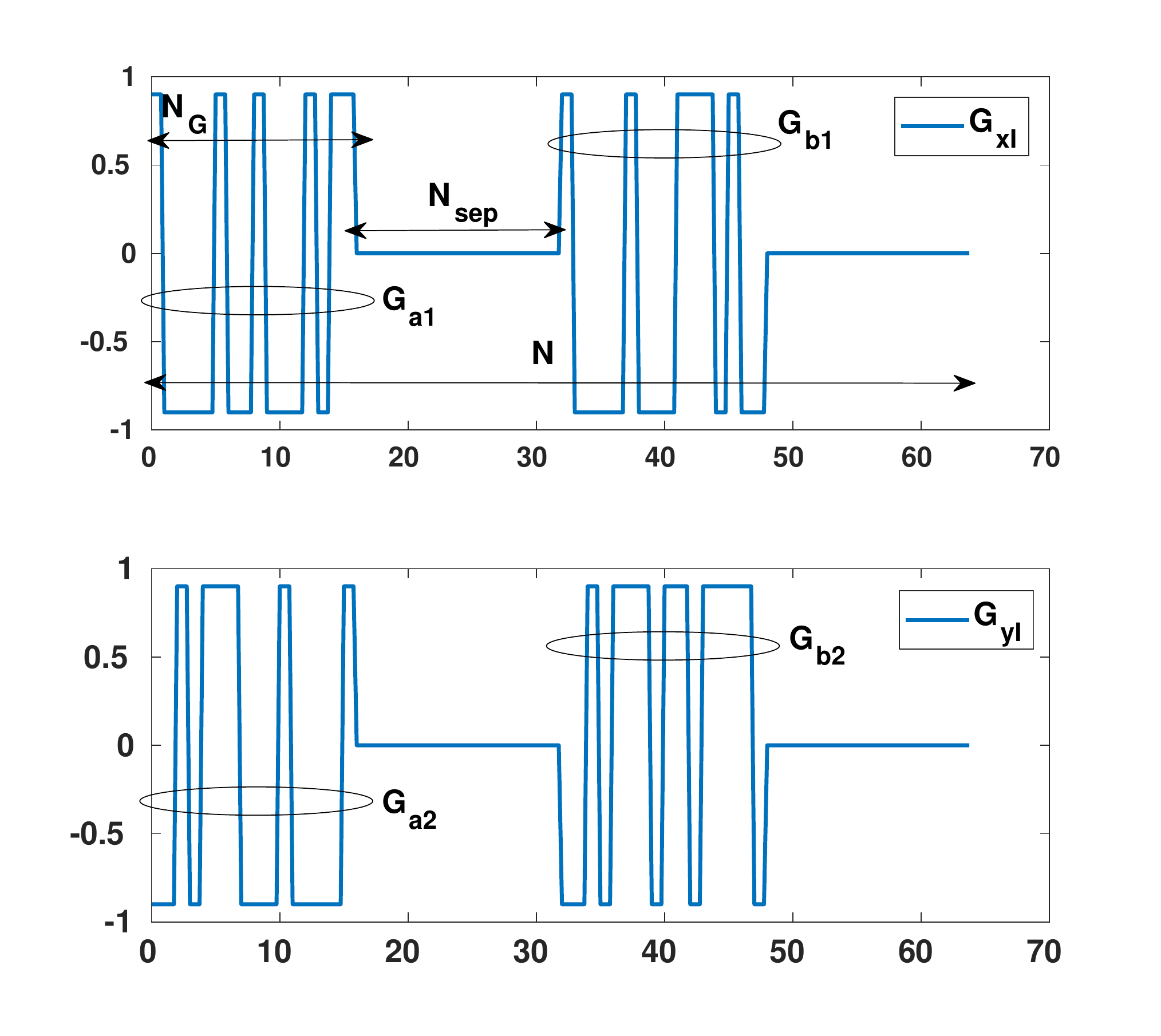}}
\put(100,0){\rotatebox{0}{\footnotesize \bf (a)}} 
}
\put(180,0)
{
\put(0,8){\includegraphics[width=220pt,height=170pt]{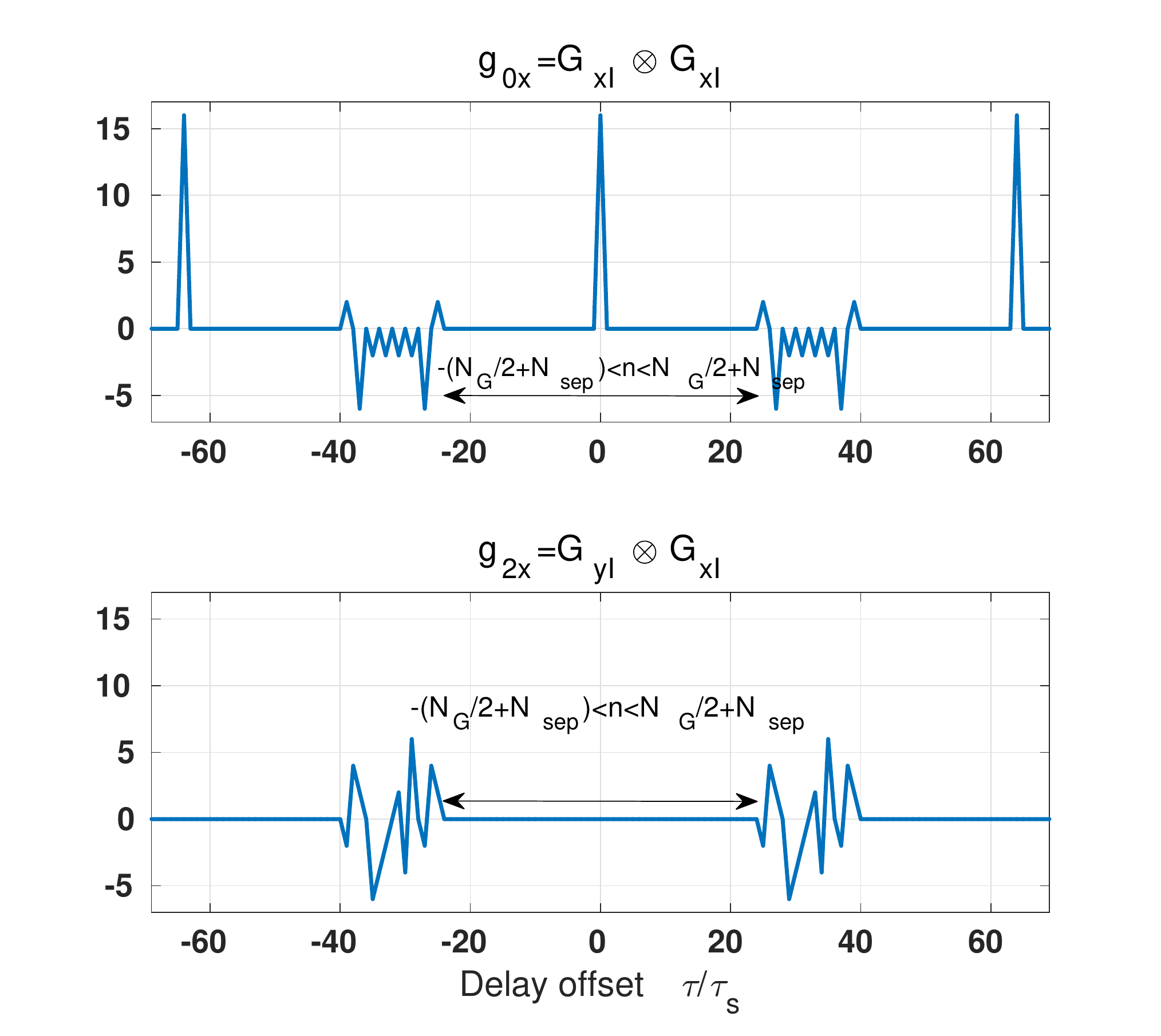}}
\put(100,0){\rotatebox{0}{\footnotesize \bf (b)}}
}
\end{picture}
\vspace{-10pt}
\caption{(a) PDM-BPSK sequences. (b) Auto- and cross- correlations with PDM-BPSK.}\label{fig:GolayBPSK}
\vspace{-5pt}
\end{figure}
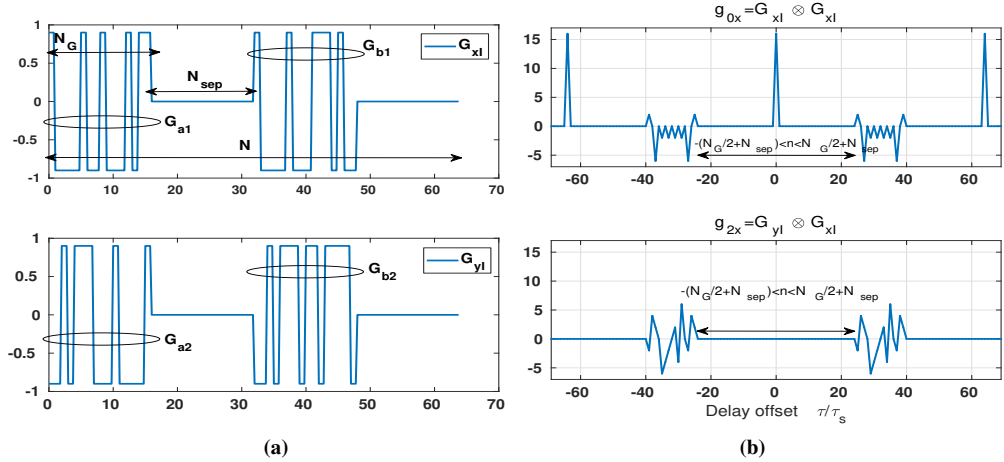

At the receiver side, a correlation is performed between the received signal $E_{rx}(n)$ and the code sent over $E_{tx}(n)$ to extract $h^\prime_{xx}(n)$ the estimate of $h_{xx}(n)$:
\begin{equation}
\begin{split}
h^\prime_{xx}(n) &= E_{rx}(n)\otimes\left(G_{xI}(n)+i G_{xQ}(n)\right)\\
&= \left( h_{xx}\ast(G_{xI}+i G_{xQ})+h_{xy}\ast(G_{yI}+i G_{yQ})\right) \otimes\left(G_{xI}+i G_{xQ}\right)\\
&= h_{xx}\ast(G_{xI}+i G_{xQ})\otimes(G_{xI}+i G_{xQ})+h_{xy}\ast(G_{yI}+i G_{yQ})\otimes(G_{xI}+i G_{xQ})\\
&= h_{xx}\ast(G_{xI}\otimes G_{xI}+ G_{xQ}\otimes G_{xQ} + i(G_{xQ}\otimes G_{xI}-G_{xI}\otimes G_{xQ})) \\
&+ h_{xy}\ast(G_{yI}\otimes G_{xI}+G_{yQ}\otimes G_{xQ} + i(G_{yQ}\otimes G_{xI}-G_{yI}\otimes G_{xQ})) \\
&= h_{xx}(n)\ast(g_{0x}(n)+ig_{1x}(n))+h_{xy}(n)\ast(g_{2x}(n)+ig_{3x}(n))
\end{split}
\label{hxx}
\end{equation}
where we partially dropped the $n$ index for clarity and define the following sequences: 
\begin{equation}
\begin{split}
g_{0x}(n) &= G_{xI}(n)\otimes G_{xI}(n) + G_{xQ}(n)\otimes G_{xQ}(n)\\
g_{1x}(n) &= G_{xQ}(n)\otimes G_{xI}(n) - G_{xI}(n)\otimes G_{xQ}(n)\\
g_{2x}(n) &= G_{yI}(n)\otimes G_{xI}(n) + G_{yQ}(n)\otimes G_{xQ}(n)\\
g_{3x}(n) &= G_{yQ}(n)\otimes G_{xI}(n) - G_{yI}(n)\otimes G_{xQ}(n)\\
\end{split}
\end{equation}
Hence, the conditions for perfect estimation of $h_{xx}(n)$, i.e. $E\left[ h^\prime_{xx}(n)\right] = h_{xx}(n)$ are:
\begin{equation}
g_{0x}(n) = \delta(n),~~g_{1x}(n) =g_{2x}(n) =g_{3x}(n) = 0
\label{eq:cd1}
\end{equation}
Similarly, $E_{rx}(n)$ is correlated with the code sent over $E_{ty}(n)$ to extract $h^\prime_{xy}(n)$: 
\begin{equation}
\begin{split}
h^\prime_{xy}(n) &= E_{rx}(n)\otimes\left(G_{yI}(n)+i G_{yQ}(n)\right)\\
&= h_{xx}\ast(G_{xI}\otimes G_{yI}+ G_{xQ}\otimes G_{yQ} + i(G_{xQ}\otimes G_{yI}-G_{xI}\otimes G_{yQ})) \\
&+ h_{xy}\ast(G_{yI}\otimes G_{yI}+G_{yQ}\otimes G_{yQ} + i(G_{yQ}\otimes G_{yI}-G_{yI}\otimes G_{yQ})) \\
&= h_{xx}(n)\ast(g_{2y}(n)+ig_{3y}(n))+h_{xy}(n)\ast(g_{0y}(n)+ig_{1y}(n))
\end{split}
\label{hxy}
\end{equation}
Again, the conditions for perfect estimation of $h_{xy}(n)$ come down to: 
\begin{equation}
g_{0y}(n) = \delta(n),~~g_{1y}(n) =g_{2y}(n) =g_{3y}(n) = 0
\label{eq:cd2}
\end{equation}
Developing the correlation equations with $E_{ry}$ instead of $E_{rx}$ to estimate $h_{yx}(n)$ and $h_{yy}(n)$ yields the same  conditions as those in Eqs.~\eqref{eq:cd1} and~\eqref{eq:cd2}.
To build polarization-multiplexed training sequences satisfying these conditions, let us consider two mutually orthogonal complementary pairs of Golay sequences $\lbrace G_{a1},G_{b1}\rbrace$ and $\lbrace G_{a2},G_{b2}\rbrace$: 
\begin{equation}
\begin{split}
G_{a1}(n)\otimes G_{a2}(n) + G_{b1}(n)\otimes G_{b2}(n) &=  0\\
G_{a1}(n)\otimes G_{b1}(n) + G_{a2}(n)\otimes G_{b2}(n) &=  0\\
\end{split}
\label{Had1}
\end{equation}
The proof of existence of mutually orthogonal pairs of complementary sequences can be found in~\cite{Hua06}. One basic example set of sequences of size $N_G=4$ satisfying these properties is: $G^4_{a1}=\left[1,-1,-1,-1\right] $, $G^4_{b1}=\left[-1,1,-1,-1\right] $, $G^4_{a2}=\left[-1,-1,1,-1\right] $, $G^4_{b2}=\left[1,1,1,-1\right] $. Larger sequences of length $N_G=2^{p+2},p\geq1$ are derived recursively:
\begin{equation}
\begin{split}
G^{N_G}_{a1} &= \left[ G^{N_G/2}_{a1}, G^{N_G/2}_{b1} \right] \\
G^{N_G}_{b1} &= \left[ G^{N_G/2}_{a1},-G^{N_G/2}_{b1} \right] \\
G^{N_G}_{a2} &= \left[ G^{N_G/2}_{a2}, G^{N_G/2}_{b2} \right] \\
G^{N_G}_{b2} &= \left[ G^{N_G/2}_{a2},-G^{N_G/2}_{b2} \right]
\end{split}
\label{Gol2}
\end{equation}
We now study the feasibility of polarization-multiplexed transmission of training sequences jointly satisfying properties~\eqref{Gol1} and~\eqref{Had1} to achieve perfect impulse response estimation of the Jones matrix, then we define the mapping of these sequences over binary modulation formats. First, we modulate one polarization channel and set the other polarization to zero. To measure $h_{xx}(n)$ and $h_{yx}(n)$, the two sequences of a single Golay pair $\lbrace G_{a1},G_{b1}\rbrace$ are transmitted successively through a binary phase shift keying (BPSK) modulation (one coded bit per symbol $\lbrace -1,1\rbrace$) on $G_{xI}(n)$ while $G_{xQ}(n)=G_{yI}(n)=G_{yQ}(n)=0$ resulting in $g_{1x}(n)=g_{2x}(n)=g_{3x}(n)=0$ and $g_{0x}(n)=G_{xI}(n)\otimes G_ {xI}(n)$. The transmitted signal can be expressed as:
\begin{equation}
E_{tx}(n+kN) = \begin{cases}
~~G_{a1}(n) & 0\leq n<N_G\\
~~0 & N_G\leq n <N_G+N_{sep}\\
~~G_{b1}(n-N_G-N_{sep}) & N_G+N_{sep}\leq n < 2N_G+N_{sep}\\
~~0 & 2N_G+N_{sep}\leq n < N
\end{cases}
\end{equation}
The code is defined as the two complementary sequences sent successively with a guard interval of length $N_{sep}$ following each sequence, as shown in the upper part of Fig.~\ref{fig:GolayBPSK}(a). The code periodicity is then $N = 2(N_G+N_{sep})$ symbols. The upper part of Fig.~\ref{fig:GolayBPSK}(b) shows the periodic auto-correlation of this code highlighting a zero-auto-correlation zone in the range $-(N_G/2+N_{sep})<n<(N_G/2+N_{sep})$. This finite zero-correlation zone translates into a constraint on the impulse response of the sensor array: to achieve perfect estimation of $h_{xx}$ and $h_{xy}$, the sensor array channel response must spread over a time $T_{IR}<(N_G/2+N_{sep})T_S$. Moreover, $N_{sep}$ can be set to 0 yielding $N=2N_G$, consequently maximizing the duty cycle. Hence, the sensor array can be continuously interrogated with a periodic code made of two complementary sequences sent successively such that $T_{code}=NT_S>4T_{IR}$.

We now extend the single polarization case to dual polarization states, applying a BPSK modulation on each of the two orthogonal polarization states (Polarization Division Multiplexing or PDM). A Golay pair $\lbrace G_{a1},G_{b1}\rbrace$ is applied to $G_{xI}(n)$ and a mutually orthogonal pair $\lbrace G_{a2},G_{b2}\rbrace$ is simultaneously applied to $G_{yI}(n)$ shown in the lower part of Fig.~\ref{fig:GolayBPSK}(a). The auto-correlation for $G_{yI}(n)$ has the same properties as for $G_{xI}(n)$. The lower part of Fig.~\ref{fig:GolayBPSK}(b) shows the cross-correlation $g_{2x}(n)$ between $G_{yI}(n)$ and $G_{xI} (n)$ expected to be null over a window of $2N_G$ samples. Therefore, a perfect estimation of the Jones matrix of the sensor array is possible using polarization-coded BPSK sequences in the same conditions as in the single polarization case.
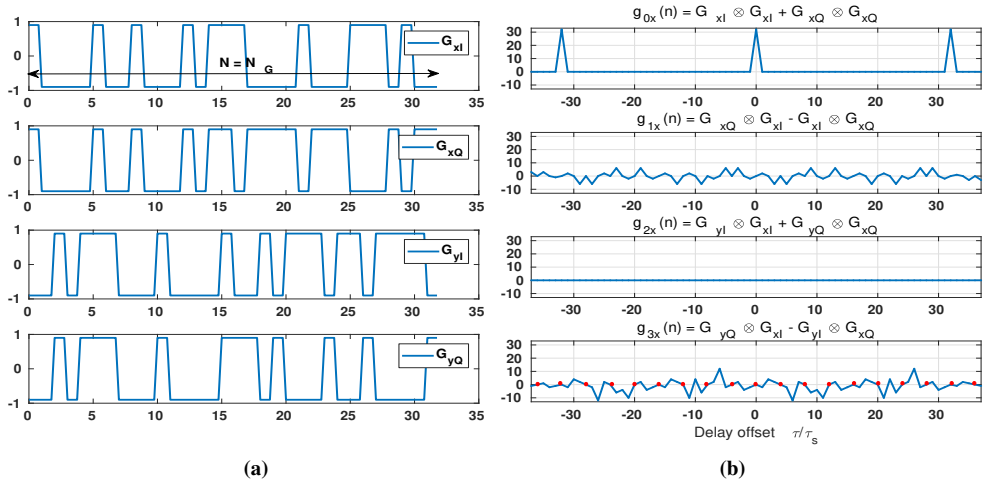
\begin{figure}[htbp]
\centering
\begin{picture}(400,180)
\put(0,0)
{
\put(-10,5){\includegraphics[width=220pt,height=180pt]{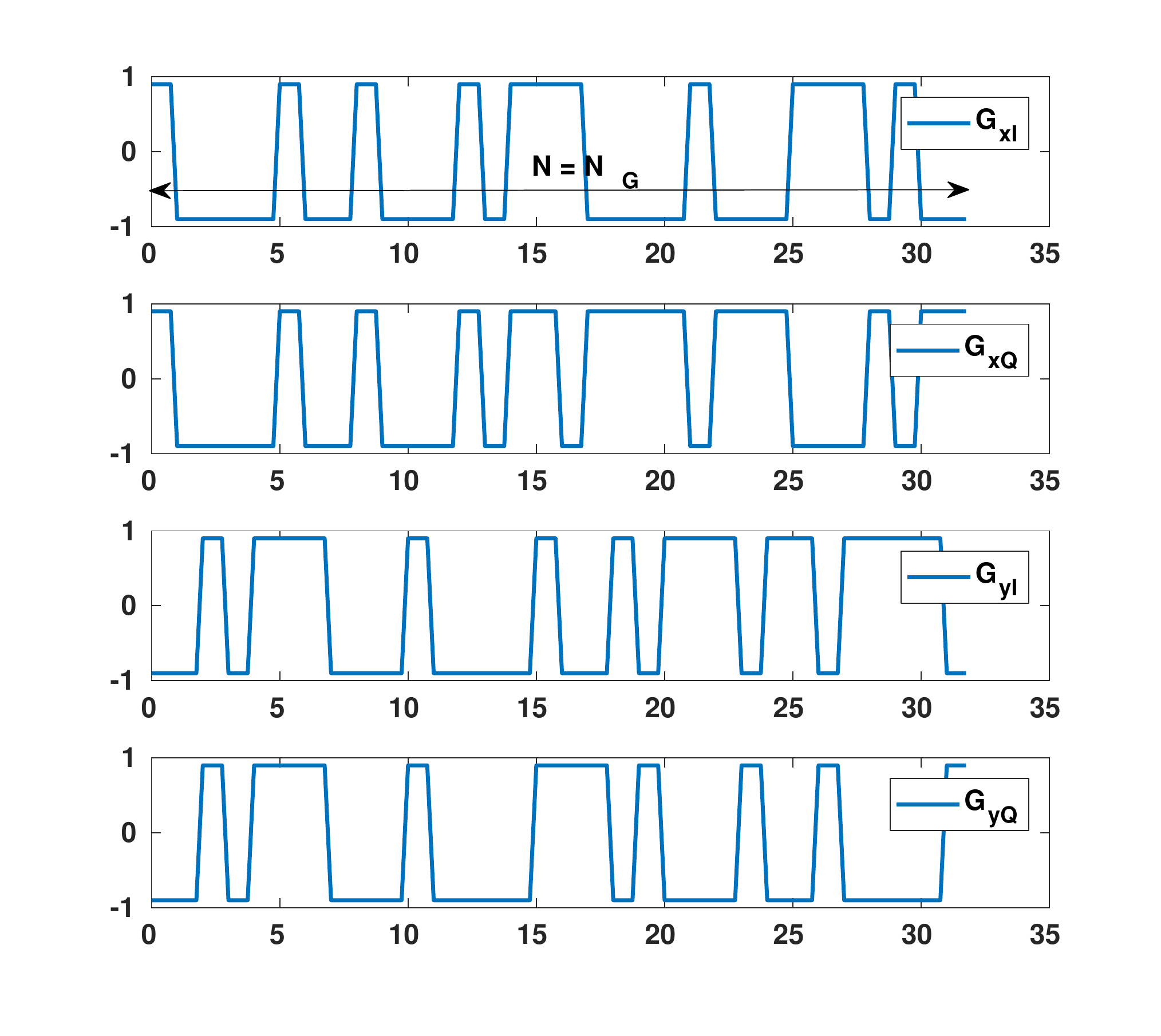}}
\put(100,0){\rotatebox{0}{\footnotesize \bf (a)}} 
}
\put(180,0)
{
\put(0,8){\includegraphics[width=220pt,height=180pt]{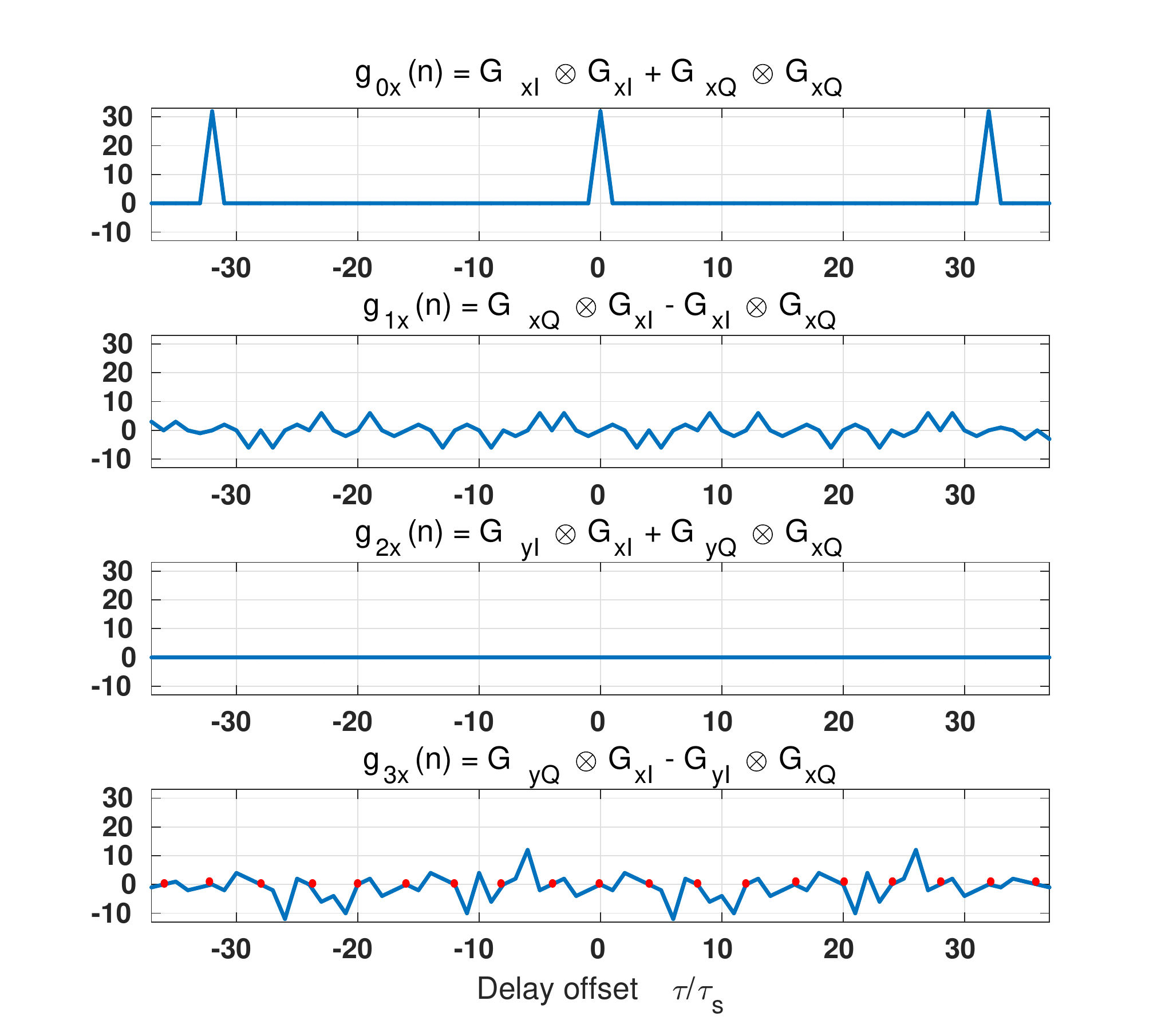}}
\put(100,0){\rotatebox{0}{\footnotesize \bf (b)}}
}
\end{picture}
\vspace{-10pt}
\caption{(a) PDM-QPSK sequences. (b) Auto- and cross- correlations with PDM-QPSK.}\label{fig:GolayQPSK}
\vspace{-5pt}
\end{figure}

Minimizing the probing time $T_{code}$ is desirable to increase the number of channel impulse response measurements per second, hence expanding the covered bandwidth. Instead of temporally multiplexing the two sequences from the complementary pair to probe the channel, we may shorten this probing time by half if we modulate the two complementary sequences in phase and quadrature over each polarization tributary through a quadrature phase shift keying (QPSK) modulation. A QPSK constellation consists of the following four complex numbers of unit energy: $\sqrt{2}/2\lbrace 1+i,-1+i,1-i,-1-i\rbrace$. Keeping $N_{sep}=0$, we set $E_{tx}(n)=G_{a1}(n)+iG_{b1}(n)$ and $E_{ty}(n)=G_{a2}(n)+iG_{b2}(n)$ to create the PDM-QPSK coded sequences with $N=N_G$, as shown in Fig.~\ref{fig:GolayQPSK}(a). Recalling the properties in Eqs.~\eqref{Gol1} and~\eqref{Had1}, we get $g_{0x}(n)=g_{0y}(n)=\delta(n)$ and $g_{2x}(n)=g_{2y}(n)=0$. Thus, Eqs.~\eqref{hxx} and~\eqref{hxy} come down to:
\begin{equation}
\begin{split}
h^\prime_{xx}(n) &= h_{xx}(n) +i(h_{xx}(n)\ast g_1(n)+h_{xy}(n)\ast g_3(n))\\
h^\prime_{xy}(n) &= h_{xy}(n) -i(h_{xy}(n)\ast g_1(n)+h_{xx}(n)\ast g_3(-n)) 
\end{split}
\end{equation}
where $g_1(n)=G_{b1}(n)\otimes G_{a1}(n)-G_{a1}(n)\otimes G_{b1}(n)$ and $g_3(n)=G_{b2}(n)\otimes G_{a1}(n)-G_{a1}(n)\otimes G_{b2}(n)$. The auto- and cross- correlation terms for this modulation scheme in Fig.~\ref{fig:GolayQPSK}(b) show that the conditions for perfect estimation of $h_{xx}(n)$ and $h_{xy}(n)$ are not fulfilled since $g_{1x}(n)$ and $g_{3x}(n)$ are not null. However, it is noteworthy to mention that $g_{1x}(n)$ equals zero for every second index and $g_{3x}(n)$ equals zero for every fourth index. Hence, if we consider a standard equally-spaced-FBG sensor array, a perfect channel estimation is achieved subject to the following condition: the symbol interval $T_S$ is set to one fourth of the dual path delay between the reflectors, which yields $F_S=1/T_S = 4p(\frac{c_f}{2d_s})$ where $d_s$ is the distance between two consecutive FBGs, $p\in\mathbb{N}^\ast$ and $c_f=c/n_g$, $n_g$ being the group refractive index of the fiber and $c$ the velocity of light. Thus, combining polarization multiplexing and the suggested QPSK coding leads, in this specific case, to a probing period reduced by a factor of four compared to a standard use of complementary sequences, which enhances the sensitivity and/or extends the bandwidth of the measurement system. In this case, the new constraint on the channel response for perfect estimation is $T_{code}>T_{IR}$.  

\subsection{Optical phase extraction from Jones matrix}
Even though we estimated the full Jones matrix of each fiber segment, we focus in this work only on the optical phase $\phi$ that can be computed as half the phase of the determinant of the dual-pass Jones matrix of each segment at the subsequent FBG reflector:
\begin{equation}
\phi = 0.5\angle(h^\prime_{xx}h^\prime_{yy}-h^\prime_{xy}h^\prime_{yx})
\end{equation} 
The phase is periodically estimated to capture its evolution at each sensor and at consecutive times, achieving a spatio-temporal map of the mechanical/acoustic events surrounding the sensor array with a spatial resolution of $d_s$ and an estimate computed each $T_{code}$ seconds. Being interested in the phase evolution in each fiber segment, the differential phase can be computed with the first reflected phase selected as a reference. 

\section{Experimental setup}
\begin{figure}[htbp]
\centering\includegraphics[width=400pt]{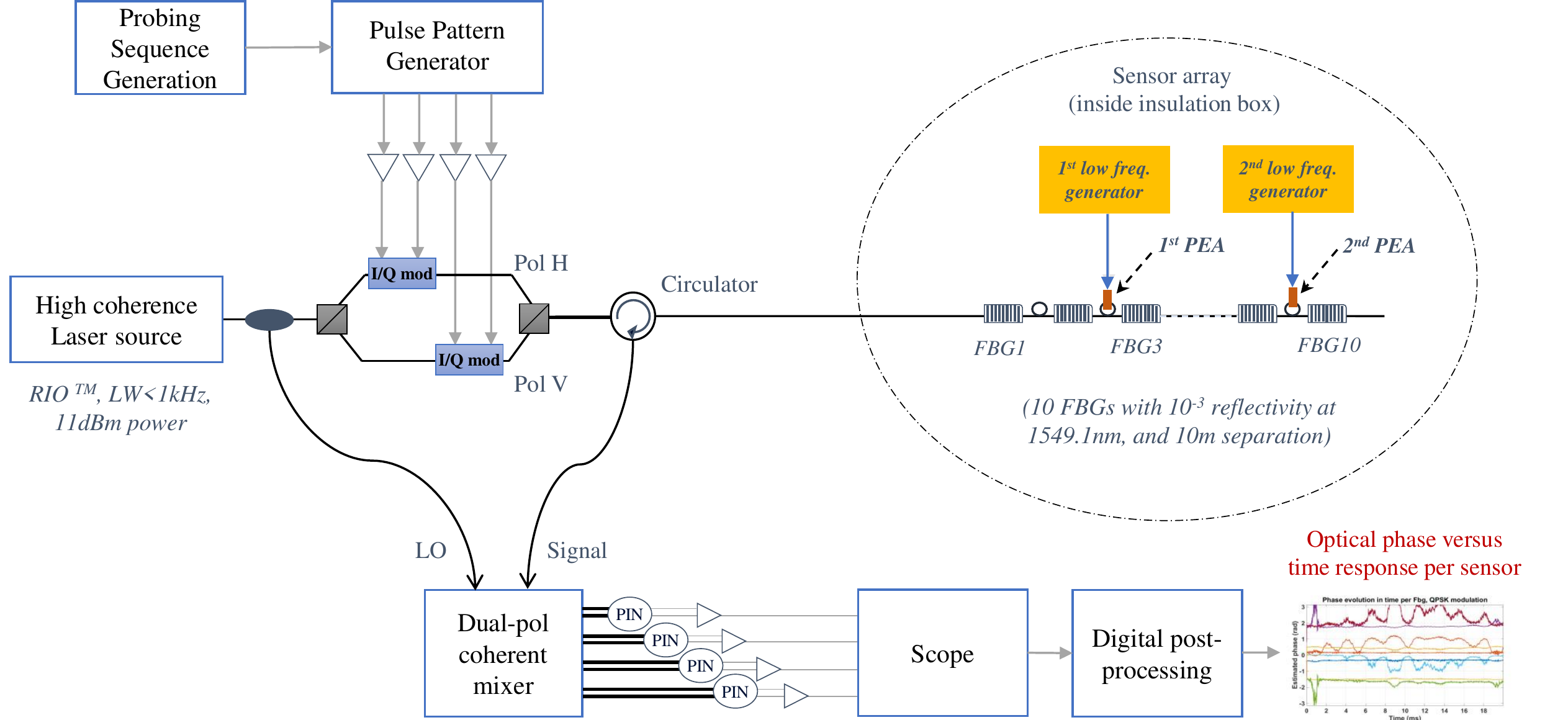}
\caption{Experimental Setup (PEA: piezoelectric actuator).}
\label{fig:ExpSetup}
\end{figure}

The experimental test-bed consists of a coherent transmitter and receiver - similar to the ones used in long-haul optical communication systems - forming the interrogator, and connected to the sensor array through an optical circulator as shown in Fig.~\ref{fig:ExpSetup}. The light from a $\mathrm{RIO}^{\mathrm{TM}}$ laser with a linewidth of $600~\textrm{Hz}$ emitting a power of $10~\mathrm{dBm}$ at $\lambda_0=1549.1~\mathrm{nm}$ is split into two to be used as a carrier at the transmitter and a local oscillator at the receiver (self-homodyne configuration). The sensor array consists of 10 FBGs with a reflectivity of $10^{-3}$  separated by $10~\mathrm{m}$ of fiber. The dual path optical delay between two $d_s$-spaced FBG reflectors is $\tau_s = 2n_gd_s/c$. The symbol duration $T_S$ has to be selected to fulfil $T_S=\tau_s/K$ where $K\geq1$ is an integer. For $d_s=10~\mathrm{m}$, the symbol rate $1/T_S$ has to be chosen as a multiple of $40~\textrm{MHz}$.

At the transmitter side, the carrier is modulated using a dual-polarization I/Q Mach-Zehnder modulator. Four RF signals accounting for the in-phase and quadrature components of each polarization are generated at various symbol rates (multiples of $40~\textrm{MHz}$) and amplified before reaching the modulator. The probing sequences are continuously generated without any guard band. The modulated optical signal is then injected in the sensor array through a circulator. 

The reflected light from the FBGs goes through a circulator towards a dual-polarization coherent mixer used to detect the in-phase and in-quadrature components over two polarization states. The different interfering optical signals are detected by balanced photodiodes of $1.6~\textrm{GHz}$ bandwidth and the four RF signals $I_X$,$Q_X$,$I_Y$,$Q_Y$ are sampled at $500~\textrm{MSa/s}$ by an oscilloscope during a measurement window $T_{acq}$.

The sensor array is inserted in a mechanically-insulated box to isolate it from the lab environment (acoustic and mechanical vibrations from lab occupants, fans of various instruments, ...). To accurately quantify the performance of our sensing system, two independent mechanical stimuli  are applied at two different locations: one between the second and third FBG (approximately at $25~\mathrm{m}$ from the circulator) and another between the ninth and tenth FBG (around $95~\mathrm{m}$ from the circulator). At these locations, $1.5~\textrm{m}$ of fiber is coiled around a cylindrical piezoelectric actuator having an outer diameter of $5~\mathrm{cm}$. The actuators are excited by frequency generators with sinusoidal tones with peak-to-peak amplitude $V_{pp1,pp2}$ and frequencies $f_{e1,e2}$.

\section{Experimental results}

\subsection{Static regime}
\begin{figure}[htbp]
\centering
\begin{picture}(400,150)
\put(0,0)
{
\put(0,8){\includegraphics[width=200pt,height=150pt]{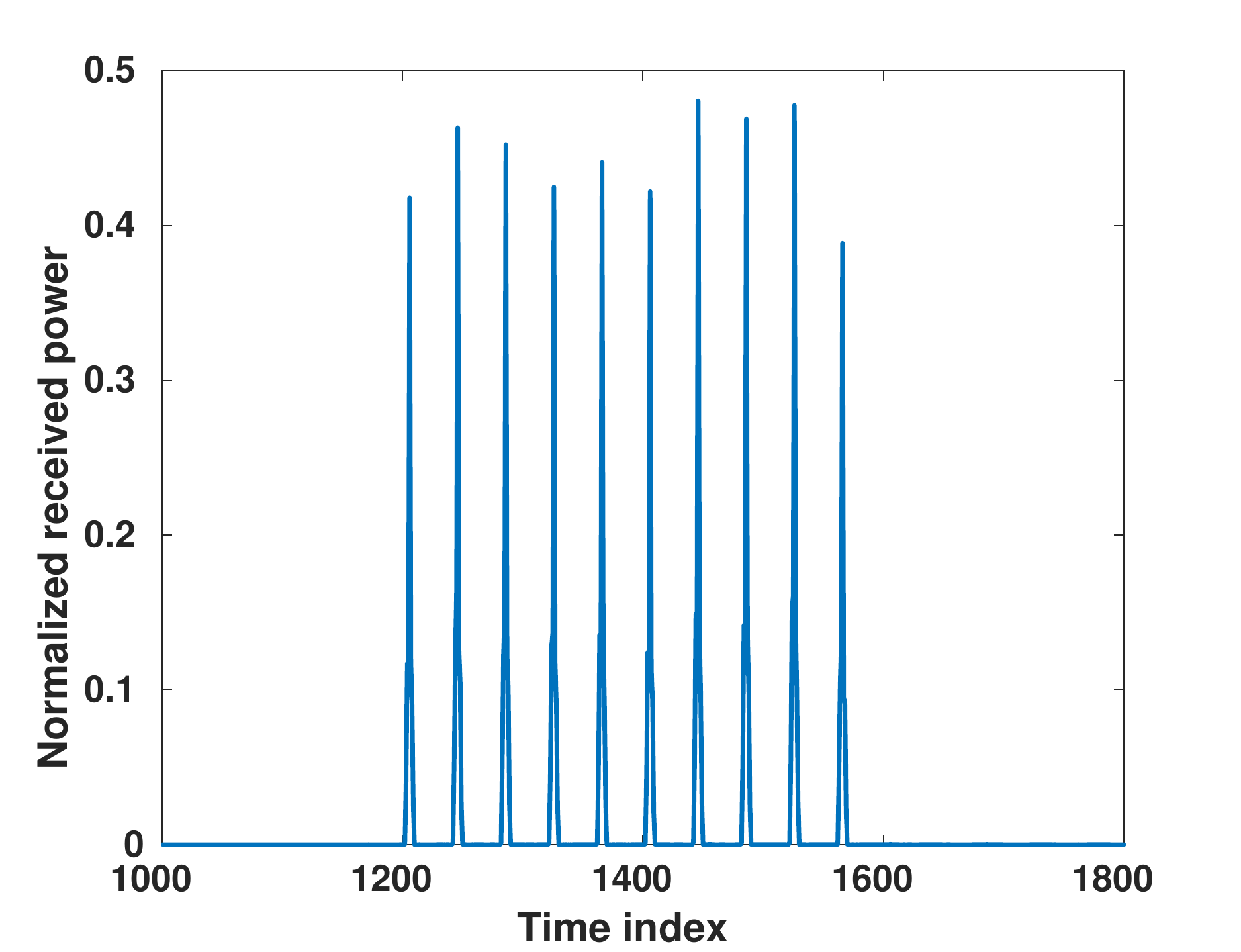}}
\put(100,0){\rotatebox{0}{\footnotesize \bf (a)}}
}
\put(200,0)
{
\put(0,8){\includegraphics[width=200pt,height=150pt]{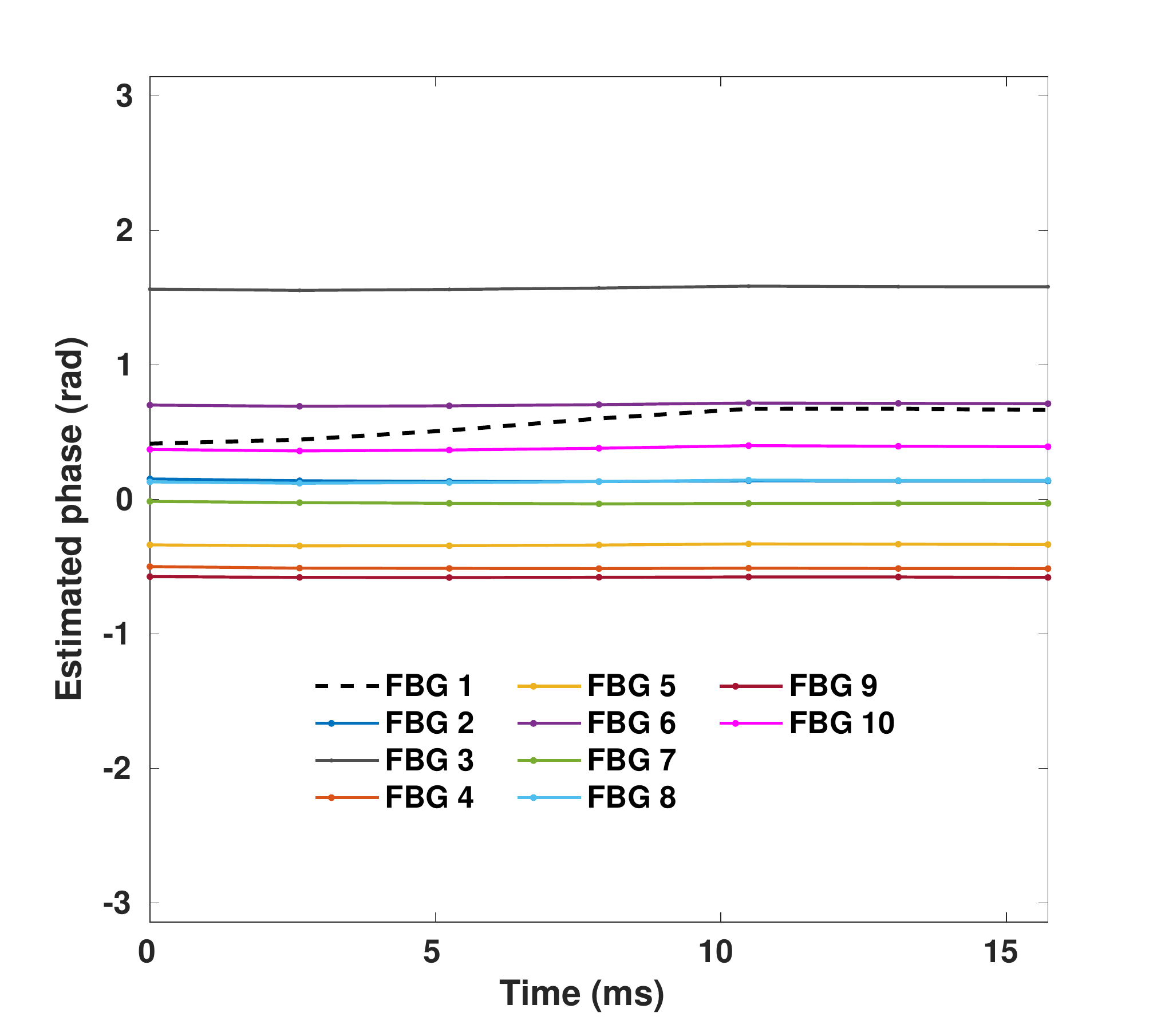}}
\put(100,0){\rotatebox{0}{\footnotesize \bf (b)}}
}
\end{picture}
\caption{(a) Measured intensities at the receiver side. (b) Estimated phases in static mode.}\label{fig:Static}
\end{figure}
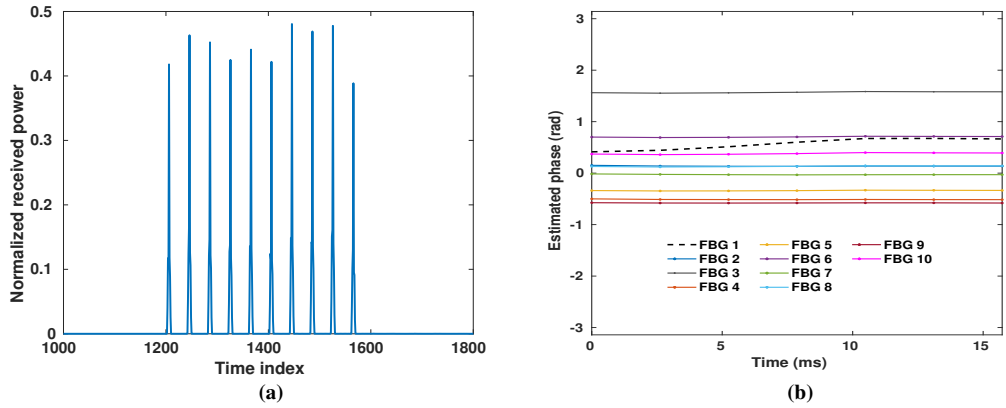

As a first step, we check the phase stability as a function of time at each FBG reflector without applying any mechanical excitation. The sensing array, placed in its insulated box, is continuously probed at a symbol rate of $160~\mathrm{MSymbol/s}$. Figure~\ref{fig:Static}(a) shows the signal intensity captured at the receiver side after a correlation process for one transmitted code. The peaks correspond to the reflections on each of the 10 FBG separated by $10~\mathrm{m}$ of fiber (corresponding to $0.1~\mathrm{\mu s}$ round trip delay). The optical phase of the signal reflected at each FBG is extracted from the Jones matrix at each peak location and the procedure is periodically repeated for each received code.

To quantify the phase stability in static mode, we measure the standard deviation (std) of the phase at each FBG over a time frame of $20~\textrm{ms}$. A capture of the estimated phases in static mode is shown in Fig.~\ref{fig:Static}(b) for the ten FBGs. Note that the first FBG serves as the optical phase reference from which the differential phase at the next FBGs is extracted; the phase of the first FBG is thus ignored in the following analyses. For each measurement, we record an average standard deviation by averaging the std values at the nine FBGs. The received signal power at the input of the coherent mixer is measured at $-27~\mathrm{dBm}$. 

\begin{figure}[htbp]
\centering
\begin{picture}(400,150)
\put(0,0)
{
\put(-5,5){\includegraphics[width=200pt,height=150pt]{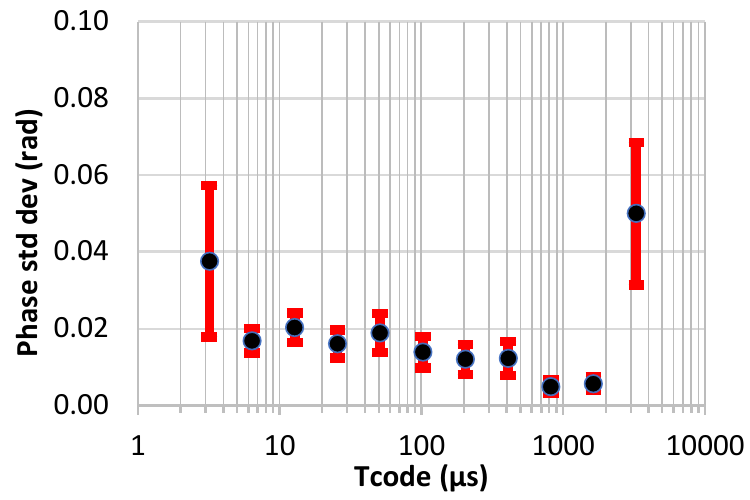}}
\put(100,0){\rotatebox{0}{\footnotesize \bf (a)}} 
}
\put(200,0)
{
\put(-5,5){\includegraphics[width=190pt,height=150pt]{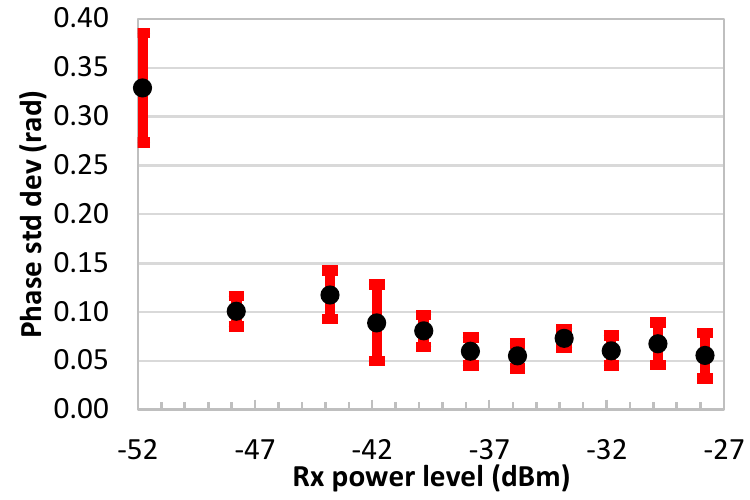}}
\put(100,0){\rotatebox{0}{\footnotesize \bf (b)}}
}
\end{picture}
\caption{(a) Standard deviation of estimated phases as a function of code length. (b) Standard deviation of estimated phases as a function of signal power at the receiver input.}\label{fig:CodeLength_SNRmargin}
\end{figure}
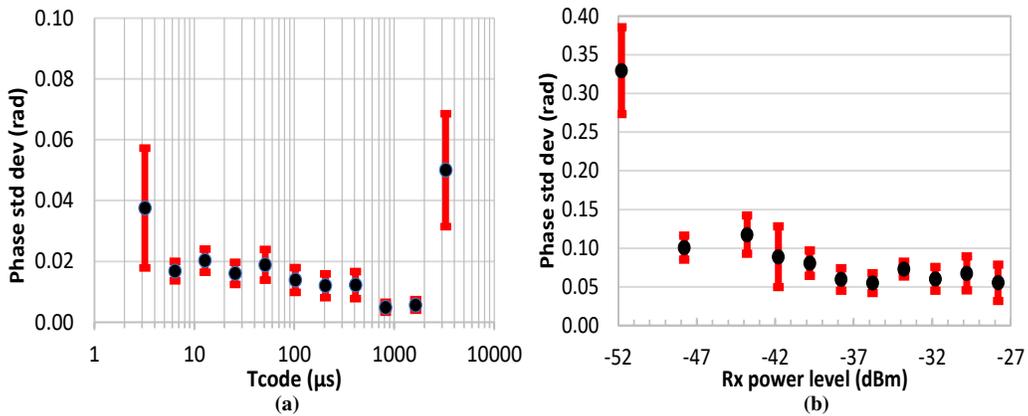

Next, we measure this average std for various lengths of the probing code. The choice of the code length is driven by the trade-off between the measurement noise and the coherence length of the laser source: when probing the sensor array with a very short code, the collected energy over a single code is low which makes us vulnerable to the receiver noise; conversely, a very long code spreads over a duration that exceeds the coherence time of the laser source which invalidates the phase reference and the relative phases computed subsequently. This is illustrated in Fig.~\ref{fig:CodeLength_SNRmargin}(a) where the phase standard deviation increases on the left edge (for codes shorter than $3~\mathrm{\mu s}$ and on the right edge (for codes longer than $3~\mathrm{ms}$) because of the correlation noise and of the coherence loss respectively. The used laser source - dedicated to sensing applications - has a $600~\mathrm{Hz}$ linewidth, corresponding to a coherence time of $0.5~\mathrm{ms}$. Between these two limits, the standard deviation of the phase is relatively constant around $10~\mathrm{mrad}$. 

Later on, we test the dependence of this result to the optical power of the signal at the input of the coherent receiver by fixing the code length to $3.2~\mathrm{\mu s}$ and varying the signal power level from $-27~\mathrm{dBm}$ down to $-52~\mathrm{dBm}$. The local oscillator power was fixed to $7~\mathrm{dBm}$. In a dual-polarization coherent receiver used in a homodyne configuration, the detected in-phase and in-quadrature photocurrents at the outputs of the balanced photodiodes $I_I$ and $I_Q$ for the two polarization states $X$ and $Y$ are given by:
\begin{equation}
\begin{split}
I_{I,X/Y} &\propto \sqrt{P_{S,X/Y}P_{LO,X/Y}}\cos(\phi_{X,Y}+\phi_{LO}) + \eta_{I,X/Y}\\
I_{Q,X/Y} &\propto \sqrt{P_{S,X/Y}P_{LO,X/Y}}\sin(\phi_{X,Y}+\phi_{LO}) + \eta_{Q,X/Y}
\end{split}
\end{equation} 
where $P_{S,X/Y}$ is the optical power of the signal projected on the $X$ (resp. on the $Y$) polarization axes of the receiver, $P_{LO,X/Y}$ is the optical power of the local oscillator projected on the $X$ (resp. on the $Y$) polarization, $\phi_{X,Y}$ stands for the optical phase of the signal projected on the $X$ (resp. on the $Y$) polarization, $\phi_{LO}$ is the laser phase noise, and $\eta_{I/Q,X/Y}$ is an additive white Gaussian noise added at the receiver side (a combination of shot noise and thermal noise). Figure~\ref{fig:CodeLength_SNRmargin}(b) shows that the phase stability slowly deteriorates when reducing the signal power down to $-48~\mathrm{dBm}$ at the receiver input. This slow decrease was investigated through numerical simulations to study the noise sources at the receiver side and was found to be mainly due to a limitation imposed by relative intensity noise (RIN) of the laser and shot noise at the photodiodes rather than thermal noise. The swift rise of the std below $-48~\mathrm{dBm}$ is due to a phase unwrapping problem when phase variations become too high.
\begin{figure}[htbp]
\centering\includegraphics[width=7cm]{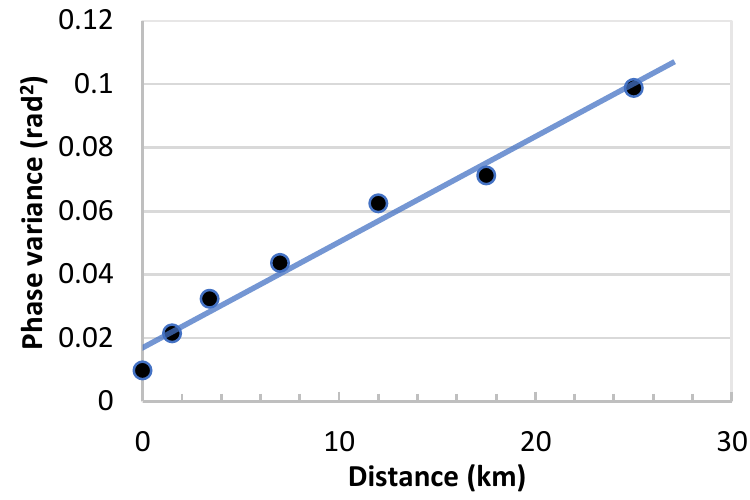}
\vspace{-5pt}
\caption{Standard deviation of estimated phases as a function of reach.}
\vspace{-5pt}
\label{fig:Reach}
\end{figure}

Another important feature of an optical fiber sensing system is its reach or the maximum distance that can be covered. In our scheme, the reach limit is given by the increase in phase noise for increased round-trip distances. To assess the reach, fiber spools of increasing lengths were added between the interrogator and the sensor array, and the average standard deviation of the estimated phases were computed for each length. Figure~\ref{fig:Reach} shows the obtained results. The used code length was $82~\mathrm{\mu s}$, and the observation window fixed to $40~\mathrm{ms}$. A tenfold degradation of the standard deviation is observed when moving from $0$ to a one-way distance of $25~\mathrm{km}$. Furthermore, beyond $34~\mathrm{km}$, the estimated phases are corrupted by a phase noise with much larger phase variance resulting from phase unwrapping errors (round-trip distance of $68~\mathrm{km}$ approaching the coherence length of the laser source (around $l_c=c_f/(\pi\Delta\nu)=100~\mathrm{km}$ for $\Delta\nu=600~\mathrm{Hz}$).

\subsection{Dynamic regime}
The sensor array is now tested in dynamic mode by means of two identical piezoelectric actuators placed at 25m and 95m from the sensor input. The used actuator is a $5~\mathrm{cm}$-outer-diameter ring with a radial efficiency of $400~\mathrm{pm/V}$. $1.5~\mathrm{m}$ of fiber is wound around each piezo, leading to $25~\mathrm{nm}$ of fiber extension per one volt of excitation voltage. We measured a phase shift of 1 radian for a $75~\mathrm{nm}$ fiber extension obtained by applying an excitation of 3 Volts.

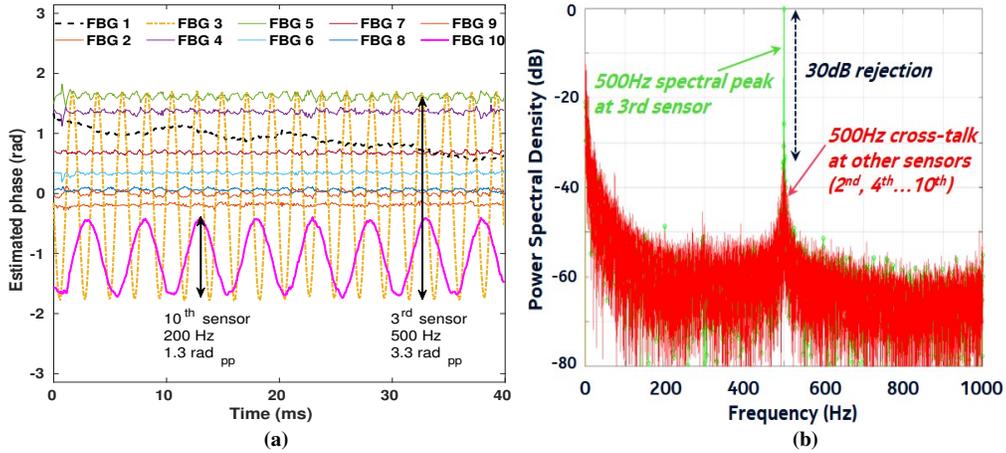
\begin{figure}[htbp]
\centering
\begin{picture}(400,170)
\put(0,0)
{
\put(-8,5){\includegraphics[width=220pt,height=175pt]{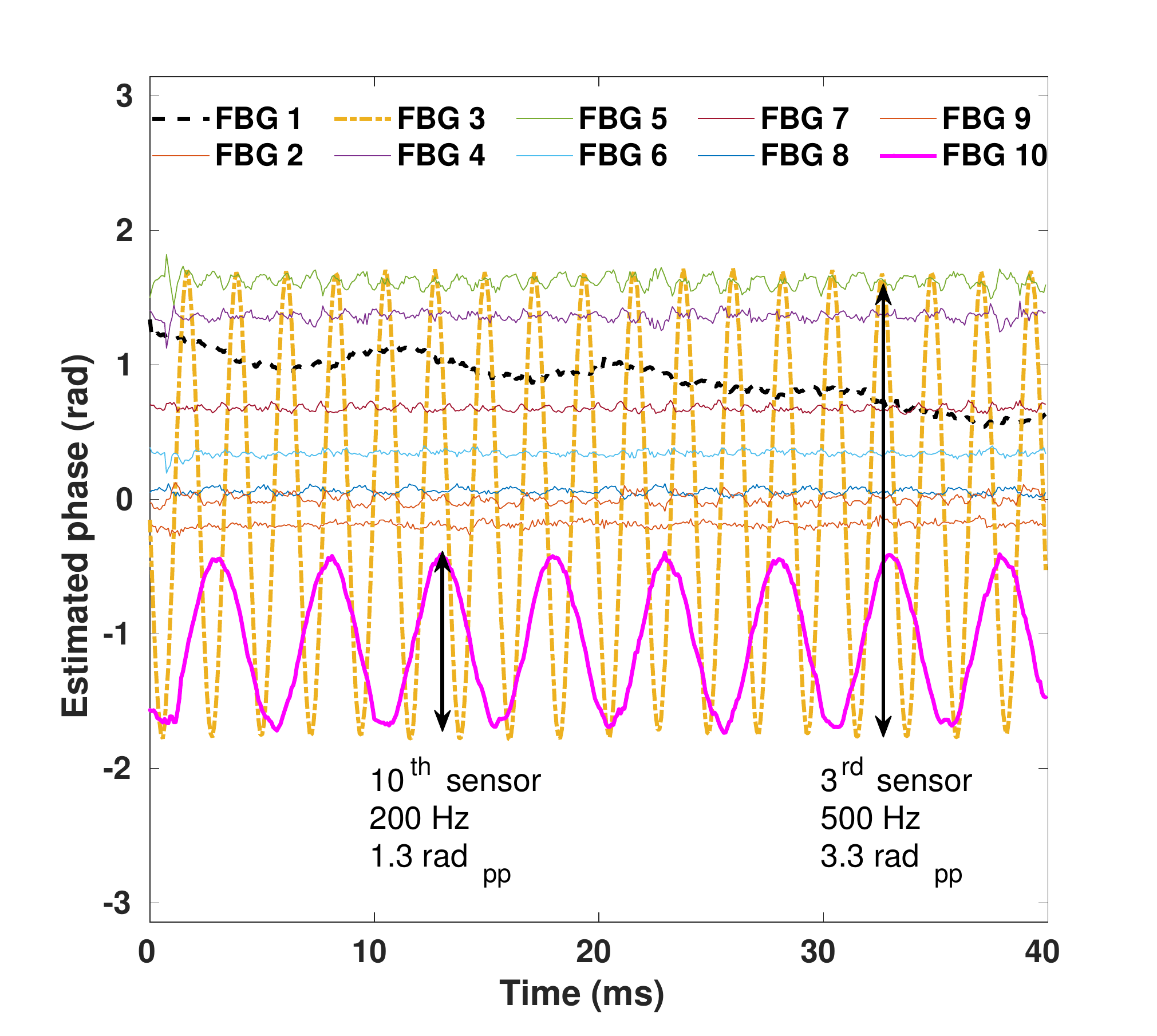}}
\put(100,0){\rotatebox{0}{\footnotesize \bf (a)}} 
}
\put(200,0)
{
\put(-5,3){\includegraphics[width=195pt,height=175pt]{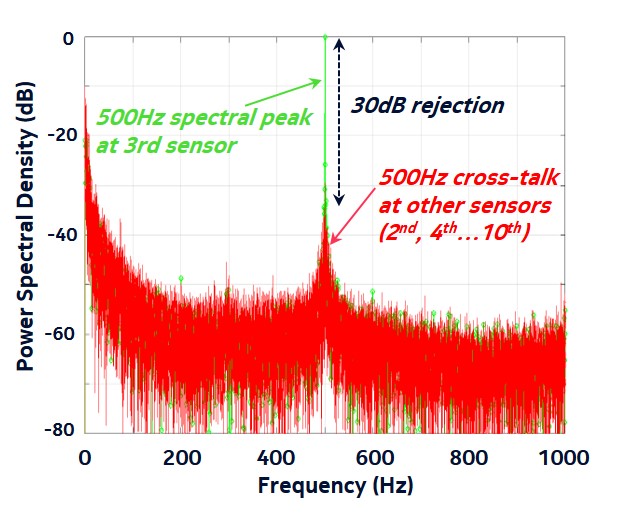}}
\put(100,0){\rotatebox{0}{\footnotesize \bf (b)}}
}
\end{picture}
\vspace{-10pt}
\caption{(a) Distributed sensing capability showing low crosstalk between sensors. (b) Crosstalk level at other sensors when only a single one is excited.}\label{fig:DisSens}
\vspace{-5pt}
\end{figure}

As a first test, we simultaneously apply a $500~\mathrm{Hz}$ (resp. $200~\mathrm{Hz}$) sine wave with a $10~\mathrm{Vpp}$  (resp. $4~\mathrm{Vpp}$) magnitude on the first (resp. second) actuator. The sensor array is probed with a $82~\mathrm{\mu s}$-long PDM-QPSK code. Figure~\ref{fig:DisSens}(a) shows the phase measured as a function of time at each of the 10 FBGs. The black dotted curve represents the absolute phase measured at the first FBG. More interesting are the phase evolutions captured at the third and tenth FBGs: both sine waves are easily identified and their magnitudes are simply scaled by the radial efficiency of the actuator. Furthermore, phases measured at the other FBG locations are stable as a function of time, proving the absence of crosstalk between sensors. We performed additional measurements with a single active actuator excited by a $500~\mathrm{Hz}$ sine wave to quantify the minimum rejection over all the remaining unexcited segments and measured a crosstalk rejection level of $-30~\mathrm{dB}$ as shown in Fig.~\ref{fig:DisSens}(b).	

\begin{figure}[htbp]
\centering
\begin{picture}(400,150)
\put(0,0)
{
\put(-5,5){\includegraphics[width=190pt,height=140pt]{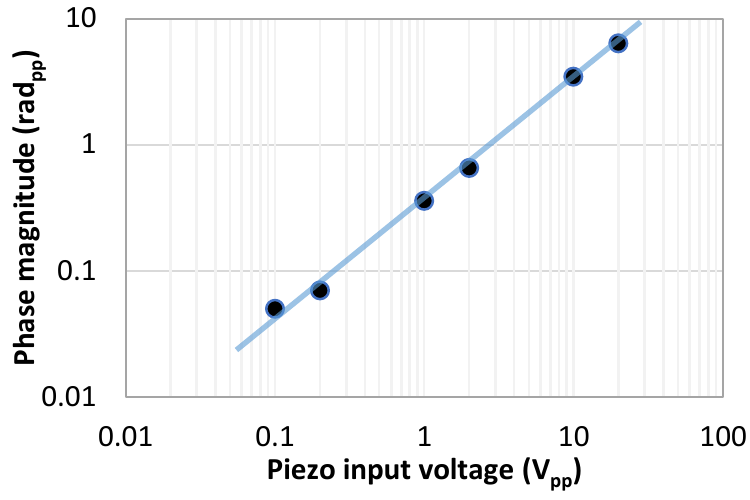}}
\put(100,0){\rotatebox{0}{\footnotesize \bf (a)}} 
}
\put(200,0)
{
\put(-10,7){\includegraphics[width=190pt,height=140pt]{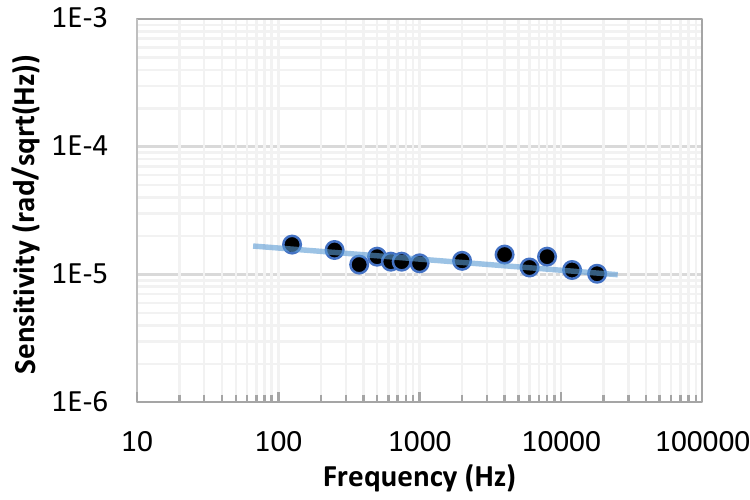}}
\put(100,0){\rotatebox{0}{\footnotesize \bf (b)}}
}
\end{picture}
\caption{(a) Dynamic range: peak-to-peak phase magnitude versus peak-to-peak voltage for a $1~\mathrm{kHz}$ sine wave. (b) Sensitivity in $\mathrm{rad/\sqrt{Hz}}$ for sine waves between $\left[100:18000\right]~\mathrm{Hz}$.}\label{fig:Dyn_Sen}
\end{figure}
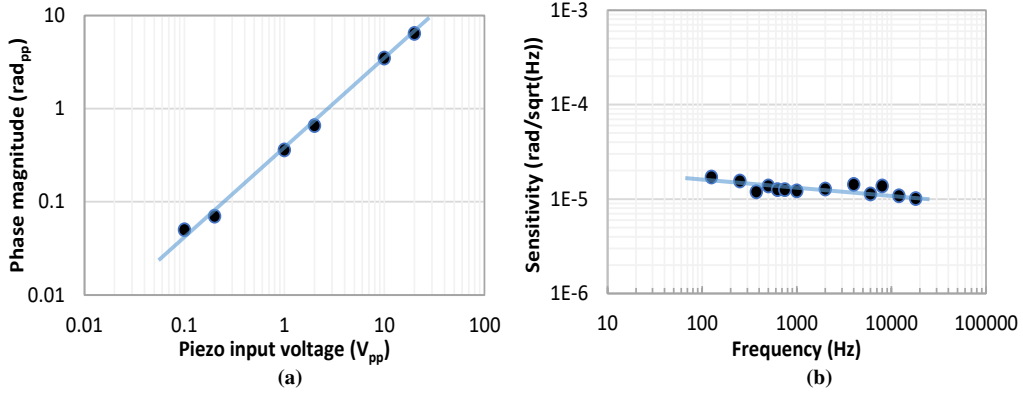

Later on, the evolution of the phase magnitude as a function of the excitation voltage has also been quantified and is shown in Fig.~\ref{fig:Dyn_Sen}(a) for a $1~\mathrm{kHz}$ sine wave. The probing code length is $82~\mathrm{\mu s}$ and the observation window is $40~\mathrm{ms}$ long. We observe a linear behavior ($20~\mathrm{dB}$ dynamic range) for voltages between $0.1~\mathrm{Vpp}$ and $20~\mathrm{Vpp}$. We could not further increase the voltage with our low-frequency signal generator. Below $0.1~\mathrm{V}$, the noise floor induced by the phase noise of the laser will become prominent. The dynamics of the sensing system can be enhanced with a laser having a narrower linewidth. However, this demonstrated dynamic range is already acceptable to analyze a wide range of mechanical signals. Next, we measure the sensitivity of our system or the smallest detectable change in the sensed variable at a given frequency, often expressed in terms of $\mathrm{rad}/\sqrt{\mathrm{Hz}}$. For that, one piezoelectric actuator is excited with a pure tone of constant amplitude producing a $2\pi$ peak-to-peak phase variation. The phase is captured over an observation window of $8~\mathrm{ms}$ and the probing code length is fixed to $20.48~\mathrm{\mu s}$. Sensitivity is computed from the normalized power spectral density of the estimated phase at the FBG following the sine wave stimulus as $\sqrt{N_{B}/F_{max}}$ where $N_{B}$ is the noise power in a frequency resolution of $B=125~\mathrm{Hz}$ corresponding to the $8~\mathrm{ms}$ observation window and $F_{max}=1/(2T_{code})$ is the maximum mechanical bandwidth of the used code ($24.4~\mathrm{kHz}$ in this case)~\cite{Pos00}. The measured sensitivity between $10~\mathrm{and}~20~\mathrm{\mu rad}/\sqrt{\mathrm{Hz}}$ is shown in Fig.~\ref{fig:Dyn_Sen}(b) for frequencies in the range of $\left[100:18000\right]~\mathrm{Hz}$. Furthermore, we measured the sensitivity after displacing the sensor array by adding $25~\mathrm{km}$ of SMF and noticed a tenfold deterioration in sensitivity (between $100~\mathrm{and}~200~\mathrm{\mu rad}/\sqrt{\mathrm{Hz}}$).

\begin{figure}[htbp]
\centering\includegraphics[width=250pt,height=210pt]{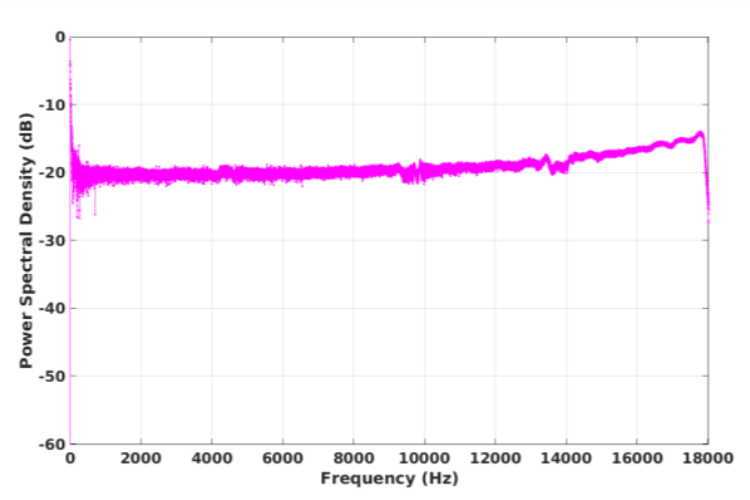}
\caption{Power spectral response of the system measured over the audio bandwidth.}
\label{fig:Linearity}
\end{figure}

Fixing the probing code length at $26~\mathrm{\mu s}$, which corresponds to a mechanical bandwidth of $19~\mathrm{kHz}$, the power spectral response is now measured by applying on the actuator a $1~\mathrm{s}$-long chirp excitation that linearly explores the audio bandwidth from $20~\mathrm{Hz}$ to $18~\mathrm{kHz}$. The obtained power spectral density of the phase response from the stimulated sensor is shown in Fig.~\ref{fig:Linearity}. The disturbances visible on the left part of the figure are induced by the limited measurement window, they disappear when measuring the low frequency part during a larger window. The rise in the response observed above $10~\mathrm{kHz}$ comes from the actuator, and is induced by its first resonance peak located at $20~\mathrm{kHz}$ (provided by the manufacturer of the piezoelectric actuator). This resonance peak can be digitally compensated, and the power spectral response would be flat within $\left[20:18000\right]~\mathrm{Hz}$. This linearity, added to the previously showcased dynamic range and sensitivity, demonstrates the ability of the system to reliably capture distributed audio/mechanical signals over a wide spectral range, as large as the human hearing system. Although the demonstrated range is bounded by $18~\mathrm{kHz}$ in this work, the use of shorter probing codes generated at higher symbol rates will allow us to explore even higher frequencies.

%

\section{Conclusion}
We introduced novel polarization-multiplexed codes derived from Golay sequences providing a perfect optical channel estimation for phase and polarization sensing applications - while covering only the former in this paper - with a code length flexibility that adjusts the system mechanical bandwidth to the application requirement. The underlying setup is derived from the one used in coherent optical telecommunication systems with the requirement of a low-phase-noise laser source used in a self-homodyne configuration. To generate the probing excitation at the transmitter side, there is no need for any acousto-optic modulators, nor digital-to-analog converters (DACs) thanks to the binary nature of the proposed sequences. In addition, the sensor array can be continuously probed by periodic codes maximizing the signal-to-noise ratio and the covered bandwidth. The main limiting parameter is the laser coherence: the duration of probing codewords and the round-trip delay in the sensor array should be within the coherence time of the laser source to guarantee a targeted sensitivity value. An FBG based sensor array excited with piezoelectric actuators was used experimentally to accurately quantify the system performance. With a $600~\textrm{Hz}$ linewidth laser source modulated by our proposed PDM-QPSK code, a sensitivity of $10~\mathrm{\mu rad}/\sqrt{\mathrm{Hz}}$ was measured for mechanical perturbations up to $18000~\mathrm{Hz}$, thus covering the entire spectral range of the human hearing system. 

\section*{Acknowledgments}
We warmly thank Ole Henrik Waagaard from Alcatel Submarine Networks Norway AS for his help in the development of the theoretical section.


\end{document}